\documentclass[12pt]{article}
\usepackage{axodraw}
\usepackage{epsfig}   
\usepackage{amssymb}

\newcommand{\lsp} {\tilde{\chi}_1^0}
\newcommand{\lsim}{\raisebox{-0.13cm}{~\shortstack{$<$ \\[-0.07cm] $\sim$}}~}
\newcommand{\gsim}{\raisebox{-0.13cm}{~\shortstack{$>$ \\[-0.07cm] $\sim$}}~}
\newcommand{\eqa} {\begin{eqnarray} }
\newcommand{\eqe} {\end{eqnarray}}
\newcommand{\beq} {\begin{equation}}
\newcommand{\eeq} {\end{equation}}

\begin{document}
\pagestyle{empty}
\begin{flushright}
April 2007
\end{flushright}
\begin{center}
{\large\sc {\bf Signals of Very High Energy Neutralinos in Future Cosmic
    Ray Detectors}}

\vspace{1cm}
{\sc Sascha Bornhauser} and {\sc Manuel Drees}

\vspace*{5mm}
{\it Physikalisches Institut, Universit\"at Bonn, Nussallee 12, D53115
  Bonn,  Germany} \\
\end{center}
\vspace*{1cm}
\begin{abstract}
  ``Top--down'' models explain the observation of ultra high energy cosmic
  rays (UHECR; $E \gsim 5 \cdot 10^{19}$ eV) through the decay of very
  massive, long--lived ``$X$ particles''. If superparticles with masses near a
  TeV exist, $X$ decays also lead to a significant flux of very energetic
  neutralinos, assumed to be the (stable or long--lived) lightest
  superparticles. There is a range of energies where neutrinos get absorbed in
  the Earth, but neutralinos can still traverse it. These neutralinos could in
  principle be detected. We calculate the detection rate in planned
  experiments such as OWL and EUSO. For bino--like neutralinos, which have
  been considered previously, we find detection rates below 1 event per
  Teraton of target and year in all cases; often the rates are much smaller.
  In contrast, if the neutralino is higgsino--like, more than ten events per
  year per Teraton might be observed, if the mass of the $X$ particle is near
  its lower bound of $\sim 10^{12}$ GeV.

\end{abstract}
\newpage
\setcounter{page}{1}
\pagestyle{plain}

\pagestyle{plain}
\section{Introduction}

The existence of ultra--high energy cosmic rays (UHECR), with $E \gsim 5 \cdot
10^{19}$ eV, remains a mystery \cite{reviews}. They have been detected by
every major cosmic ray experiment, but we do not know of any astronomical
objects that has sufficiently strong electromagnetic fields extending over a
sufficiently large volume to accelerate charged particles to the required
energies. Nor do we understand how these particles, once created, can reach
us, given their energy loss through scattering on the cosmic microwave
background \cite{gzk}. 

One radical idea \cite{topdown} is that UHECR originate from the decay of very
massive, yet long--lived $X$ particles. Since one starts with very energetic
particles, which lose energy first through parton showering and fragmentation,
and later while propagating through the universe, these class of models are
known as ``top--down'' models. The most energetic CR event that has been
observed to date has $E \simeq 3 \cdot 10^{20}$ eV \cite{fly}. This implies a
lower bound $M_X \gsim 10^{12}$ GeV on the mass of the $X$ particles. Since
UHECR are observed today, the lifetime of $X$ must be at least comparable to
the age of the Universe. Several particle physics models containing candidates
with sufficiently large mass and long lifetime have been suggested
\cite{reviews,models}. Ways to produce these particles in the very early
universe are discussed in \cite{topdown,prodx}.

Models of this type can be made compatible with all existing data, including
the first data from the Pierre Auger observatory \cite{auger_top}. However, in
order to decisively test these models, one has to find predictions that allow
to discriminate between top--down and the more conventional bottom--up
\cite{reviews} models. These two classes of models usually predict somewhat
different spectra for photons and neutrinos at high energies, and/or different
distributions of the arrival directions. However, distinguishing between UHE
photons and protons is nontrivial. Gigaton class neutrino telescopes now under
construction should see some very energetic neutrinos if these models are
correct \cite{neutx}; however, bottom--up models generically also lead to
comparable neutrino fluxes. Anisotropies in the arrival direction can be
expected \cite{ani}, if $X$ particles are distributed like (or even form the)
Dark Matter in our galaxy; however, quantitative details depend on the
distribution of matter near the galactic center, which is not well
understood.

These difficulties motivate the analysis of signals where bottom--up and
top--down models make {\em qualitatively} different predictions. This may be
possible if we postulate the existence of superparticles \cite{susyrev} at or
near the electroweak energy scale. This assumption is quite natural in the
given context, since supersymmetry is the only known way to stabilize the
large hierarchy between $M_X$ and the electroweak scale against radiative
corrections.\footnote{Note that ``large'' extra dimensions do not help here,
  since by construction the ``fundamental scale'' must be at least $M_X$ in
  order to explain the observed UHECR; this is independent of the
  dimensionality of spacetime.} Since $M_X$ is much larger than the sparticle
mass scale, $X$ decays will produce large number of superparticles. This is
true even if the primary decay of $X$ only involves Standard Model (SM)
particles; in this case superparticles will be produced in the subsequent
parton shower \cite{xdec,cyrille}. All these superparticles will decay into
lightest superparticles (LSPs), assumed to be the lightest neutralino. In
contrast, bottom--up models will produce a miniscule flux of superparticles.
The reason is that the vast majority of UHE proton or photon interactions
with matter only produces additional light particles (in particular, light
mesons and baryons); the cross section for producing superparticles remains
very small even at these energies.

This raises the question how one might observe these very energetic
neutralinos. The crucial observation \cite{bdhh2} is that there is a range of
energies where neutrinos get absorbed in the Earth, whereas neutralinos can
traverse it with little or no loss of energy. The reason for this difference
is the smaller neutralino--nucleon scattering cross section, and/or the
smaller neutralino energy loss per interaction \cite{bod1}. Note that
neutralino interactions always lead to a superparticle in the final state,
which will decay back into a neutralino. An interaction will therefore not
change the total neutralino flux, but will shift it to lower energies, where
it is (even) more difficult to detect.

In this article we provide a detailed calculation of the neutralino event
rates that one might expect in future cosmic ray detectors with very large
target volumes, like OWL \cite{owl} and EUSO \cite{euso}. We improve on
existing analyses \cite{bdhh2,luis,mele} in several ways. We use neutralino
spectra impinging on Earth calculated with the most complete code for $X$
particle decays \cite{cyrille}, where we analyze several different primary
decay modes. We also carefully include the effects of neutralino propagation
through the Earth, using the results of \cite{bod1}. Our calculation of the
event rates includes a cut on the visible energy deposited by a neutralino
interaction; since this interaction again produces an invisible neutralino,
the visible energy is usually significantly smaller than the energy of the
incoming neutralino. Moreover, we investigate both bino-- and higgsino--like
neutralinos; the cross sections for the latter have also been computed in
\cite{bod1}. We find that higgsino--like neutralinos would in fact be much
easier to detect; bino--like neutralinos most likely remain out of reach even
for the planned EUSO and OWL missions. Finally, we calculate the neutrino
background from the same model of $X$ decays as the signal.

The remainder of this article is organized as follows. The calculation of the
event rates is described in Sec.~2. In Sec.~3 we present numerical results,
and Sec.~4 is devoted to a brief summary and some conclusions.

\section{Calculation of Event Rates}

Neutralinos are produced along with protons, photons, electrons and neutrinos
at the location of $X$ decays, following a prolonged parton shower
\cite{xdec,cyrille}. We fix the normalization through the proton flux at
$10^{20}$ eV, which we take to be
\beq \label{e1}
E^3 F_p(E) = 1.6 \cdot 10^{24} \ {\rm eV}^2 {\rm m}^{-2} {\rm s}^{-1} {\rm
  sr}^{-1} 
\eeq
at $E = 10^{20}$ eV. This roughly corresponds to the flux observed by the
HiReS experiment \cite{hires}, which is somewhat smaller than that observed by
AGASA \cite{agasa}. Note, however, that we ignore the contribution of photons
to the UHECR flux. This is phenomenologically motivated by the observation
that UHECR events seem to be proton--like, rather than photon--like
\cite{prot}. Normalizing to the sum of the proton and photon fluxes would
obviously reduce the predicted neutralino flux, and hence the event rate;
depending on the $X$ decay model, the reduction factor would roughly lie
between two and five. On the other hand, we ignore all propagation effects. If
most $X$ decays occur at significant distance from our galaxy, which may well
be true if $X$ particles are confined to topological defects, both the proton
and photon fluxes might be depleted by propagation, while leaving the
neutralino (and neutrino) flux essentially unchanged. The presence of
significant propagation effects would therefore increase the predicted
neutralino flux on Earth.

Neutralinos can interact with nucleons either through the exchange of a squark
in the $s-$channel, or through the exchange of a $Z^0$ or $W^\pm$ gauge boson
in the $t-$channel. In the following we treat these two contributions, which
essentially do not interfere \cite{bod1}, in turn, before discussing the
calculation of the neutrino--induced background.

As explained in \cite{luis,mele,bod1}, the {\em $s-$channel} contribution is
dominated by the exchange of on--shell squarks. The event rate is given by:
\beq \label{nsch}
{\cal N}_s = \sum_q \int_{E_{\rm min}}^{E_{\rm max}} dE_{\rm vis} \int_{X_{\rm
           min}}^{X_{\rm max}} dX 
           \int_0^{y_{\rm max_q}} dy \frac {1}{y} 
           F_{\lsp}(\frac {E_{\rm vis}}{y},X)
           \frac {d\sigma_{s}(\frac {E_{\rm vis},}{y},y)} {dy} {\mathcal V}
           \, .
\eeq
Here, $F_{\lsp}$ is the differential neutralino flux, which depends on the
neutralino energy as well as the matter depth\footnote{Matter depth $X$ is
  costumarily given as a column depth, measured in g/cm$^2$ or, in natural
  units, in GeV$^3$; for the Earth, $X \in [0, \, 2.398 \cdot 10^6$ GeV$^3]$
  \cite{GQRS96}.} $X$. The sum runs about all quark flavors $q$, and the first
integration is over the visible energy $E_{\rm vis}=E_{\tilde \chi^0_{1,{\rm
      in}}}-E_{\tilde \chi^0_{1,{\rm out}}}=yE_{\tilde \chi^0_{1,\rm in}}$.
The factor $1/y$ appears because we integrate over the visible, rather than
total, energy. The lower limit $E_{\rm min}$ on $E_{\rm vis}$ is determined by
the energy sensitivity of the experiment, whereas the upper limit $E_{\rm
  max}$ is determined by kinematics, $E_{\rm max} \sim M_X/2$; however, after
propagation through the Earth the neutralino flux at the highest kinematically
allowed energy is very small. The lower bound on the column depth, $X_{\rm
  min}=0.13 \cdot 10^6$ GeV$^3$, corresponds to an angular cut of about $5\%$
on the signal, i.e. we only count events that emerge at least five degrees
below the horizon; this cut greatly reduces the neutrino background. $X_{\rm
  max}=2.398 \cdot 10^6$ GeV$^3$ is the maximal earth column depth,
corresponding to neutralinos that emerge vertically out of the Earth.  The
kinematic maximum of the scaling variable $y$, for 2--body decays $\tilde q
\rightarrow q + \lsp$, is $y_{\rm max_q} = 1 - m_{\lsp}^2 / m_{\tilde q}^2$.
Since the maximal neutralino energy is finite, there should strictly speaking
also be a non--vanishing lower bound on $y$; note that we need the neutralino
flux at $E_{\lsp} = E_{\rm vis}/y$. An explicit expression for the
differential cross section $d \sigma_s / d y$ can be found in \cite{bod1}.
Finally, the constant factor ${\mathcal V}$ is given by
\beq \label{vfactor}
{\mathcal V} \equiv 2\pi V_{\rm eff} \epsilon_{\rm DC} t N_A \rho_w J_D \, .
\eeq
Here, $V_{\rm eff}$ is the water equivalent (w.e.) effective volume,
$\epsilon_{DC}$ is the duty cycle (the fraction of time where the experiment
can observe events), $t$ is the observation time, $N_A = 6.022 \times 10^{23}
\mbox{~g}^{-1}$ is Avogadro's number, $\rho_w = 10^6 \mbox{~g} \mbox{m}^{-3}$
is the density of water, and $J_D = \mid\!d\cos\theta/dX\!\mid$ is the
Jacobian for the transformation $\cos\theta \rightarrow X(\cos\theta)$.

The {\em $t-$channel} exchange diagrams predominantly lead to the production
of heavier neutralinos or charginos in the final state \cite{bod1}, which we
collectively denote by $\tilde \chi_{\rm out}$. The visible energy therefore
also depends on the $\tilde \chi_{\rm out}$ decay kinematics. The event rate
can be written as:
\beq \label{ntch}
{\cal N}_t = \int_{E_{\rm min}}^{E_{\rm max}} dE_{\rm vis} 
             \int_{X_{\rm min}}^{X_{\rm max}} dX 
           \int_0^1 dy \frac {1}{y} F_{\lsp} (\frac {E_{\rm vis}} {y}, X) 
      \left( G_{\lsp}^{\rm NC}(E_{\rm vis},y) + G_{\lsp}^{\rm CC}(E_{\rm
           vis},y)\right) {\mathcal V} \, .
\eeq
Here we have written the contributions from charged and neutral currents
separately. Each term is given by a convolution of a differential cross
section for the production of $\tilde \chi_{\rm out}$ with the $\tilde
\chi_{\rm out}$ decay spectrum. These convolutions are more easily written in
terms of the variable $z = E_{\tilde \chi^0_{1,{\rm out}}} / E_{\tilde
  \chi^0_{1,{\rm in}}} = 1 - y$:
\beq \label{Gs}
G_{\lsp}^{NC,CC}(E_{\rm vis},y) =
\int_z^{z_{1,{\rm max}}} \frac{d\!z_1}{z_1}
\frac {d\sigma^{NC,CC}_{t_{\tilde \chi}} (\frac {E_{\rm vis}} {y},z_1)}{dz_1}
\left. \frac {1} {\Gamma} \frac {d\Gamma_{\tilde \chi_{\rm out}}(z_1\frac
    {E_{\rm vis}} {y}, z_2 = \frac{z}{z_1})}{dz_2} \theta(z-z_{\rm min})
  \theta(z_{\rm max} - z) \right|_{z = 1-y} \, . 
\eeq
Here $z_1 = E_{\tilde \chi_{\rm out}} / E_{\tilde \chi^0_{1,{\rm in}}}$
describes the energy transfer from the incoming lightest neutralino to the
heavier neutralino or chargino, and $z_2 = E_{\tilde \chi^0_{1,{\rm out}}} /
E_{\tilde \chi_{\rm out}}$ describes the energy transfer from this heavier
neutralino or chargino to the lightest neutralino produced in its decay. $z_2$
is chosen such that $z \equiv z_1 z_2 = 1 - y$. Explicit expressions for the
differential cross sections, and for the limits $z_{\rm min,max}, \ z_{1,{\rm
    max}}$ in Eq.(\ref{Gs}), can again be found in \cite{bod1}.\footnote{Note
  that the $G_{\lsp}^{NC,CC}$ of Eq.(\ref{Gs}) are the integration kernels
  $K_{\lsp}^{NC,CC}$ of ref.\cite{bod1}, multiplied with the total cross
  section for $t-$channel scattering.} In principle one would need to include
sums over $\tilde\chi_{\rm out}$ in Eq.(\ref{Gs}). In practice, however, a
single neutralino and a single chargino dominate neutral and charged current
reactions, respectively \cite{bod1}.

The event rates (\ref{nsch}) and (\ref{ntch}) depend on the neutralino flux
{\em after} propagation through the Earth. Of course, the propagation effects
also depend on whether $s-$ or $t-$channel exchange is dominant. We treat
these effects through straightforward numerical integration of the transport
equations, as described in \cite{bod1}.

The {\em background} is dominated by $\nu_\tau$ scattering through $t-$channel
exchange of $W$ or $Z$ bosons. At the relevant energies electron and muon
neutrinos get absorbed efficiently in the Earth. However, since $\nu_\tau$
interactions regenerate another $\nu_\tau$, albeit at lower energy, $\tau$
neutrinos can always traverse the Earth, although their energy may be reduced
drastically. Again treating charged and neutral current processes separately,
the background rate can be written as
\beq \label{back1}
{\cal N}_{\nu} = \int_{E_{\rm min}}^{E_{\rm max}} dE_{\rm vis}
                  \int_{X_{\rm min}}^{X_{\rm max}} dX 
           \int_0^1 dy \frac {1}{y} F_{\nu}(\frac {E_{\rm vis}}{y},X) 
           \left(\frac {d\sigma_{t_{\nu}}^{\rm NC}(\frac {E_{\rm vis}}{y},y)}
           {dy} + N_{\nu}^{\rm CC}(E_{\rm vis},y)\right) {\mathcal V} \, ,
\eeq
where $y = 1 - E_{\nu,{\rm in}} / E_{\nu,{\rm out}}$. In the case of NC
scattering ($Z-$exchange) the entire visible energy results from the hadronic
vertex. In case of CC scattering ($W-$exchange) we add the visible energy
released in $\tau$ decay to that produced at the hadronic vertex:
\begin{eqnarray} \label{back2}
N_\nu^{CC}(E_{\rm vis},y) &=&   \left.
\int_z^{z_{1,{\rm max}}} \frac{d\!z_1}{z_1}
\frac{d\sigma^{CC}_{\nu} (\frac {E_{\rm vis}}{y},z_1)} {dz_1}
\cdot\frac {1} {\Gamma} \frac {d\Gamma(z_1\frac {E_{\rm vis}}{y},
  z_2 = \frac{z}{z_1})}{dz_2} \theta(z-z_{\rm min}) \theta(z_{\rm max} - z)
\right|_{z = 1-y} \, . 
\nonumber \\ &&
\end{eqnarray}
This expression is formally very similar to Eq.(\ref{Gs}), which also includes
contributions to the visible energy from the decay of an unstable particle.
This treatment is conservative since it ignores the fact that a $\tau$
produced inside the target volume may decay outside of it. Moreover, if $\tau$
production and decay both occur inside the target volume, it may be possible
to use this ``double bang'' signature to remove these background events. On
the other hand, we ignore the background from $\tau$s produced outside the
target which decay inside the observed volume. This contribution should be
smaller, since one would need higher neutrino energy to produce a given
visible energy in this manner. Note that at the energies in question, $\tau$
energy losses in rock or water are no longer negligible; this reduces the
energy released in $\tau$ decay even further. Recall that after propagation
through the earth the $\nu_\tau$ flux is a steeply falling function of energy.

The background rate (\ref{back1}) is proportional to the tau neutrino flux
$F_\nu$ emerging from the Earth. The $\nu_\tau$ flux at the location of $X$
decay is usually quite small \cite{cyrille}. However, due to near--maximal
neutrino flavor mixing, the three neutrino fluxes impinging on Earth are very
nearly equal, i.e. we take one third of the total neutrino flux, normalized
according to Eq.(\ref{e1}), as estimate of the incoming $\nu_\tau$ flux. 

As mentioned above, tau neutrinos may lose much of their energy while
traversing the Earth. We solve the corresponding transport equations using the
methods of ref.\cite{bod1}. Since we are interested in very high energies, the
tau leptons produced in CC $\nu_\tau$ reactions may lose a significant
fraction of their energy before decaying. We therefore modified the standard
treatment \cite{GQRS96} in order to at least crudely estimate the effects of
$\tau$ energy loss in matter. We do this by formally treating this energy loss
as additional scattering. To this end, we modify the integration kernel in the
transport equation for $\nu_\tau$ as follows:
\beq \label{etau0}
\frac {1} {\sigma(E_y)}\frac {d\sigma(E_y,z)} {dz} \rightarrow \left.
\int \frac {1} {\sigma(E_y)} \frac {d\sigma(E_y,z_1)} {dz_1} \frac {1} {L} 
\frac {dL(z_1E_y,E^{\prime\prime})} {d E^{\prime\prime}}
dE^{\prime\prime}\right|_{z = E^{\prime\prime}/E} \, . 
\eeq
Here $E_y = E / (1-y)$ is the energy of the incident neutrino that gives rise
to a neutrino with energy $E$ after the scattering, and the function
$dL(E_{\tau,{\rm in}}, E_{\tau,{\rm out}}) / d E_{\tau,{\rm out}}$ describes
the $\tau$ energy loss. We make the very simple ansatz \cite{DHR05}
\beq \label{etau1}
\frac {d E_\tau} {d z} = - \beta \rho E_\tau \ \ \ {\rm with} \ \beta = 0.85
\cdot 10^{-6} {\rm cm^2 g^{-1}} = {\rm const.} 
\eeq
This implies $E_\tau(z) = E_\tau(0) {\rm e}^{- \beta \rho z}$. We assume that
all $\tau$s decay after traveling a distance $z_{\rm dec} = E_\tau c \tau_\tau
/ m_\tau$, where $\tau_\tau$ is the lifetime of the $\tau$ lepton and $c$
is the speed of light. Note that we estimate the average decay length from the
$\tau$ energy {\em after} propagation. This underestimates the decay length,
and hence the effect of $\tau$ energy loss. On the other hand, for
$E_{\nu_\tau} < 10^{10}$ GeV the ansatz (\ref{etau1}) overestimates the energy
loss \cite{DHR05}. Our approximation of a fixed decay length leads to
\beq \label{etau2}
\frac {dL(E^{\prime},E^{\prime\prime})} {d E^{\prime\prime}} =
\delta \left( E^{\prime\prime} - E^{\prime}\exp(-\kappa E^{\prime\prime})
\right) \, ,
\eeq
with constant $\kappa = \beta \rho c \tau_\tau / m_\tau$. The integral over
$dL / d E''$, which appears in Eq.(\ref{etau0}), is then given by:
\beq \label{etau3}
L = \int dE^{\prime\prime} \delta (E^{\prime\prime} - E^{\prime}
\exp(-\kappa E^{\prime\prime})) = \frac {1} {1 + \kappa E^{\prime}
  \exp(-\kappa E^{\prime\prime})}  \, ,
\eeq
where in the last expression $E''$ has to be interpreted as a function of
$E'$, as determined by the argument of the $\delta-$function. We can then
evaluate the integral in Eq.(\ref{etau0}): 
\beq \label{etau4}
\frac {1} {\sigma(E_y)}\frac {d\sigma(E_y,z)} {dz} \rightarrow \left.
\left( 1 + \kappa z_1 E_y \right) \exp(\kappa z_1 E_y) \frac {1} {\sigma(E_y)}
\frac {d\sigma(E_y,z^{\prime})} {dz^{\prime}}
\right|_{z^{\prime} = z_1\exp(\kappa z_1E_y)} \, .
\eeq

The obvious advantage of our simplified treatment is that it does not
necessitate the numerical evaluation of additional integrals. This would have
been very costly, since the length scales involved in $\tau$ energy loss and
decay (a few km for $E_\tau \sim 10^8$ GeV) are very much shorter than the
$\nu_\tau$ interaction length in rock ($\sim 10^3$ km for $E_{\nu_\tau} =
10^8$ GeV) \cite{DHR05}. A more accurate treatment would therefore have
required to use many more steps in $X$ when integrating the transport
equation; even with out simple treatment, or indeed without including the
effects of $\tau$ energy loss, calculating the $\nu_\tau$ flux emerging from
Earth takes up to several CPU days. On the other hand, our simplified
treatment can only give us an indication of the size of effects due to $\tau$
energy losses. We find that the effect on the $\nu_\tau$ flux emerging from
Earth is essentially negligible for $E_{\nu_\tau} \lsim 10^7$ GeV. This is
also true for $X \gsim 0.3 X_{\rm max}$, since then the flux at $E_{\nu_\tau}
> 10^7$ GeV is negligible even if the $\tau$ energy loss is ignored. However,
it can reduce the $\nu_\tau$ flux by a factor of two or more at large
$E_{\nu_\tau}$ and small $X$.

\section{Results}

We are now ready to present numerical results. Earlier estimates
\cite{bdhh2,luis} have shown that one will need at least teraton scale targets
in order to detect hadronic interactions of neutralinos in top--down models.
Currently the only technology that might allow to monitor such large targets
is optical observation from space \cite{owl,euso}. Here one detects the light,
either from Cerenkov radiation or from fluorescence, emitted by very energetic
showers in the atmosphere. The target is therefore quite thin: the neutralinos
would have to interact either in the atmosphere itself, or just below it. One
usually estimates an effective target thickness of 10 to 20 m w.e.. A teraton
target then results if one can monitor ${\cal O}(10^6)$ km$^2$ simultaneously,
which might be possible \cite{owl,euso}. One drawback of this approach is that
observations of this kind are only feasible on clear, moonless nights, leading
to a duty cycle $\epsilon_{DC}$ in Eq.(\ref{vfactor}) of only about 10\%. In
our numerical results we therefore take a target mass of 1Tt, $\epsilon_{DC} =
0.1$, and assume an observation time of one year.

As shown in \cite{bdhh2}, the expected neutralino flux depends quite strongly
on $M_X$ as well as on the dominant $X$ decay mode. Top--down models predict
rather hard spectra, i.e. $E^3$ times the flux increases with energy. Fixing
the (proton) flux at $E = 10^{20}$ eV therefore leads to smaller fluxes at $E
< 10^{20}$ eV as $M_X$ is increased. Moreover, if $M_X$ is not far from its
lower bound of $\sim 10^{12}$ GeV, much of the relevant neutralino flux is
produced early in the parton cascade triggered by $X$ decay, which is quite
sensitive to the primary $X$ decay mode. In contrast, if $M_X \gg 10^{12}$
GeV, in the relevant energy range most LSPs originate quite late in the
cascade; in that case the LSP spectrum is largely determined by the dynamics
of the cascade itself, which only depends on Standard Model interactions, and
is not very sensitive to the primary $X$ decay mode(s).

Following ref.\cite{bdhh2} we therefore study scenarios with $M_X = 10^{12}$
and $10^{16}$ GeV, for four different primary $X$ decay modes. In contrast to
previous analyses \cite{bdhh2,luis,mele} we calculate the event rates for both
bino--like and higgsino--like neutralinos. As explained in ref.\cite{bod1} the
former interact with hadronic matter almost exclusively through $s-$channel
scattering, while the latter dominantly interact through $t-$channel diagrams.

Finally, we present results for two different values of the minimal visible
energy $E_{\rm min}$. Events with visible energy as ``low'' as $10^6$ GeV
might be observable via the Cerenkov light emitted by particles in the
atmosphere with velocities exceeding the speed of light in air. On the other
hand, the fluorescence signal (observed e.g. by the HiReS experiment
\cite{hires}) can probably only be seen for energies $\gsim 10^9$ GeV. In all
cases we require the event to come from an angle at least five degrees below
the horizon. This greatly reduces the neutrino--induced background, as
explained earlier.

\begin{table}[h!] 
\begin{center}
\begin{tabular}{|c||c|c|} 
\hline
\multicolumn{3}{|c|}{{\bf Event rates for higgsino--like $\lsp$}}\\
\hline
\hline
\hline
$E_{\rm vis}\ge 10^6$ GeV, $M_X=10^{12}$ GeV & $N_{{\tilde \chi}^0_1}$ &
$N_{\nu_{\tau}}$ \\  
\hline
\hline
$q\bar q$ & $0.56$ & $0.44$ \\
\hline
$q\tilde q$ & $1.77$ & $0.79$ \\
\hline
$l\tilde l$ & $25.19$ & $1.59$ \\
\hline
$5\times q\tilde q$ & $14.84$ & $5.03$ \\
\hline
\hline
$E_{\rm vis}\ge 10^9$ GeV, $M_X=10^{12}$ GeV & $N_{{\tilde \chi}^0_1}$ &
$N_{\nu_{\tau}}$ \\  
\hline
\hline
$q\bar q$ & $0.00883$ & $0.00001$ \\
\hline
$q\tilde q$ & $0.08676$ & $0.00001$ \\
\hline
$l\tilde l$ & $4.09828$ & $0.00002$ \\
\hline
$5\times q\tilde q$ & $0.17046$ & $0.00005$ \\
\hline
\hline
\hline
$E_{\rm vis}\ge 10^6$ GeV, $M_X=10^{16}$ GeV & $N_{{\tilde \chi}^0_1}$ &
$N_{\nu_{\tau}}$ \\  
\hline
\hline
$q\bar q$ & $0.033$ & $0.050$ \\
\hline
$q\tilde q$ & $0.024$ & $0.035$ \\
\hline
$l\tilde l$ & $0.022$ & $0.033$ \\
\hline
$5\times q\tilde q$ & $0.038$ & $0.055$ \\
\hline
\hline
$E_{\rm vis}\ge 10^9$ GeV, $M_X=10^{16}$ GeV & $N_{{\tilde \chi}^0_1}$ &
$N_{\nu_{\tau}}$ \\  
\hline
\hline
$q\bar q$ & $0.003187$ & $0.000004$ \\
\hline
$q\tilde q$ & $0.002685$ & $0.000003$ \\
\hline
$l\tilde l$ & $0.006532$ & $0.000003$ \\
\hline
$5\times q\tilde q$ & $0.003668$ & $0.000003$ \\
\hline

\end{tabular}
\caption{\label{tab1} Predicted events rates per teraton and year (with duty
  cycle $\epsilon_{DC}=0.1$) for the scenario H2 of \cite{bod1}, where $\lsp$
  is higgsino--like, and for the $\nu_\tau$ induced background. Both signal
  and background depend on the mass $M_X$ of the progenitor particle, as well
  as on the primary $X$ decay mode. We show results for $X$ decays into a
  first generation quark antiquark pair (``$q \bar q$''), into a first
  generation quark squark pair (``$q \tilde q$''), into a first generation
  lepton slepton pair (``$l \tilde l$''), and into five quarks and five
  squarks (``$5 \times q \tilde q$''). We only include events that emerge from
  an angle at least five degrees below the horizon. }
\end{center}
\end{table}

We present results for higgsino-- and bino--like neutralinos in Tables 1 and
2, respectively. We saw in ref.\cite{bod1} that the cross section for
neutralino--nucleon scattering depends only weakly on details of the sparticle
spectrum if $\lsp$ is higgsino--like. In Table 1 we therefore only show
results for one scenario with higgsino--like LSP. It has an LSP mass of 300
GeV, with the second neutralino and first chargino, which are produced
predominantly in NC and CC scattering respectively, having masses of 310 and
303 GeV, respectively; the near--degeneracy of these three states is a
consequence of these states all being higgsino--like, which in turn follows if
the LSP is a rather pure higgsino state.

As expected, we find much higher event rates for $M_X = 10^{12}$ GeV than for
$M_X = 10^{16}$ GeV. In the former case we also see that the predicted event
rate depends significantly on the primary $X$ decay mode, again as
expected. The decay into a lepton plus a slepton turns out to be most
favorable. The reason is that this decay mode leads to a rather small number
of protons produced per $X$ decay, or, put differently, to a large ratio of
the LSP and proton fluxes \cite{cyrille}. Since we normalize to the proton
flux, this then leads to a rather large LSP flux. This decay mode also leads
to the hardest $\lsp$ spectrum. Since the primary $X$ decay only involves
weakly interacting (s)particles, parton showering carries away a relatively
small fraction of the energy of the original particles. The original slepton
will then eventually decay into a very energetic neutralino. As a result,
increasing the cut on $E_{\rm vis}$ by three orders of magnitude only reduces
the predicted event rate by a factor of $\sim 5$ in this case.

The second most favorable primary $X$ decay mode is the one into five quarks
and five squarks. Since we produce ten strongly interacting (s)particles
already in the very first step, each of which initiates an extended QCD
shower, the final multiplicity is very large, but the fluxes are relatively
soft. One then again needs a rather large normalization factor to reproduce
the desired proton flux (\ref{e1}) at $E = 10^{11}$ GeV. Since the $\lsp$
spectrum is quite soft, increasing $E_{\min}$ from $10^6$ to $10^9$ GeV now
reduces the predicted signal by nearly two orders of magnitude.

The worst case is $X$ decay into SM quarks only. This gives a relatively hard
proton spectrum. Moreover, superparticles are now only produced in the parton
shower. This gives a small ratio of $\lsp$ to proton fluxes, and a relatively
soft $\lsp$ spectrum. The fourth primary $X$ decay we considered, into a quark
and a squark, also leads to a relatively hard proton flux. However, since a
superparticle is produced in the primary $X$ decay, the $\lsp$ flux is larger,
and significantly harder, than for $X \rightarrow q \bar q$ decays.

We see that at least three of the four cases might lead to observable signals
if $M_X$ is near its lower bound, and if visible energies around $10^6$ GeV
can be detected. Of course, at that energy one expects a huge number of
ordinary CR induced events, $\sim 1$ event per km$^2$ and second or (including
the duty cycle) $\sim 3 \cdot 10^{11}$ events per year in an experiment
observing $10^5$ km$^2$, as required for a teraton--scale target mass
\cite{pdg}. One will therefore need an excellent discrimination against such
down--going events in order to extract the signal of at best a handful events
per year. To that end one may need to sharpen the angular cut somewhat. This
may also be desired to further reduce the $\nu_\tau$ induced background, which
in this case is within an order of magnitude of the signal. Fig.~1 shows that
for $E_{\rm min} = 10^6$ GeV, imposing a stronger angular cut will not reduce
the signal very much. This is in accord with the results of ref.\cite{bod1},
which show large neutralino propagation effects only for LSP energies well
beyond $10^7$ GeV in this case. Note, however, that typically $E_{\rm vis}
\lsim 0.1 E_{\lsp,{\rm in}}$ for higgsino--like neutralino.

\begin{figure}[h!] 
\begin{center}
\includegraphics[width=14cm]{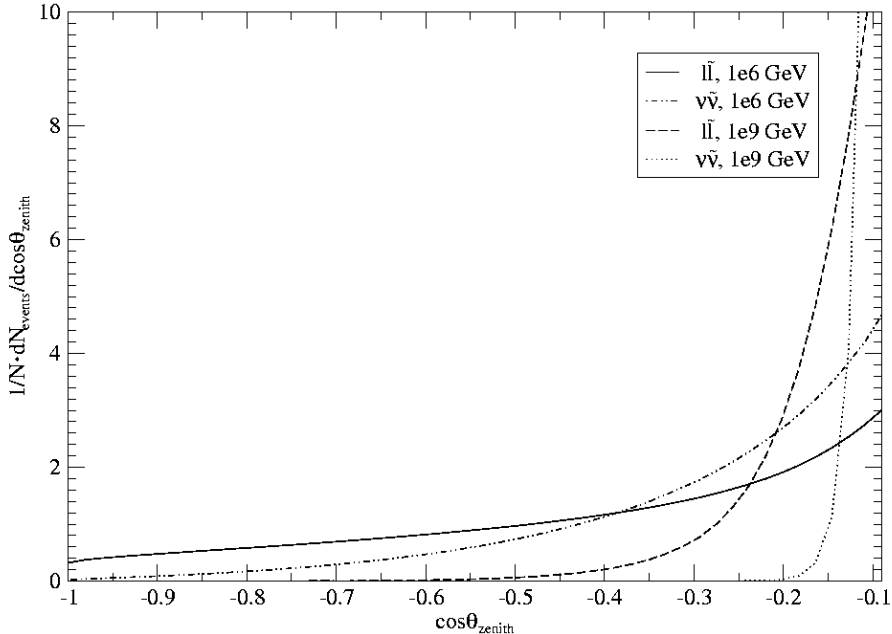}
\caption{Angular dependence of the signal from higgsino--like neutralinos from
  primary $X \rightarrow l \tilde l$ decays, and of the $\nu_\tau$ induced
  background, for two different values of the lower limit on the visible
  energy.}
\label{fig1}
\end{center}
\end{figure}

On the other hand, only the most favorable scenario remains observable if
$E_{\rm min}$ has to be increased to $10^9$ GeV. On the positive side, the
$\nu_\tau$ induced background is now at least three orders of magnitude
smaller than the signal, illustrating that the Earth can indeed be used as a
filter.  This is fortunate, since Fig.~1 shows that now the angular cut can be
sharpened only at the cost of a significant reduction of the signal. However,
in most cases one would need tens of Tt$\cdot$yr to see a convincing signal
even for $M_X = 10^{12}$ GeV; for $M_X = 10^{16}$ GeV and $E_{\rm min} = 10^9$
GeV, one would need Pt$\cdot$yr of target mass times observation time! This
would require monitoring virtually the entire surface of the Earth. The
neutralino flux from decays of such very heavy $X$ particle would remain
invisible to teraton scale detectors even for a threshold energy of $10^6$
GeV. Note that in this case the predicted event rate is almost independent of
the primary $X$ decay mode. The reason is that now the entire relevant energy
range satisfies $x \equiv 2 E / M_X \ll 1$, where the spectrum is determined
almost uniquely by the dynamics of the parton shower \cite{cyrille}.

\begin{table}[ht!] 
\begin{center}
\begin{tabular}{|c||c|c|c|} 
\hline
\multicolumn{4}{|c|}{{\bf Event rates for bino--like $\lsp$}}\\
\hline
\hline
\hline
$E_{\rm vis}\ge 10^6$ GeV, $M_X=10^{12}$ GeV & $N_{D1}$ & $N_{D2}$ & $N_{D3}$
\\  
\hline
\hline
$q\bar q$ & $0.0191$ & $0.0192$ & $0.0118$ \\
\hline
$q\tilde q$ & $0.0471$ & $0.0528$ & $0.0388$ \\
\hline
$l\tilde l$ & $0.3560$ & $0.5376$ & $0.5543$ \\
\hline
$5\times q\tilde q$ & $0.4567$ & $0.4779$ & $0.3051$ \\
\hline
\hline
$E_{\rm vis}\ge 10^9$ GeV, $M_X=10^{12}$ GeV & $N_{D1}$ & $N_{D2}$ & $N_{D3}$
\\  
\hline
\hline
$q\bar q$ & $0.00007$ & $0.00070$ & $0.00143$ \\
\hline
$q\tilde q$ & $0.00030$ & $0.00314$ & $0.00701$ \\
\hline
$l\tilde l$ & $0.00567$ & $0.06121$ & $0.14800$ \\
\hline
$5\times q\tilde q$ & $0.00201$ & $0.01982$ & $0.03967$ \\
\hline
\hline
\hline
$E_{\rm vis } \ge 10^6$ GeV, $M_X=10^{16}$ GeV & $N_{D1}$ & $N_{D2}$ &$N_{D3}$
\\
\hline
\hline
$q\bar q$ & $0.00095$ & $0.00103$ & $0.00075$ \\
\hline
$q\tilde q$ & $0.00070$ & $0.00077$ & $0.00055$ \\
\hline
$l\tilde l$ & $0.00079$ & $0.00117$ & $0.00062$ \\
\hline
$5\times q\tilde q$ & $0.00113$ & $0.00122$ & $0.00088$ \\
\hline
\hline
$E_{\rm vis} \ge 10^9$ GeV, $M_X=10^{16}$ GeV & $N_{D1}$ & $N_{D2}$ & $N_{D3}$
\\
\hline
\hline
$q\bar q$ & $0.000006$ & $0.000058$ & $0.000140$ \\
\hline
$q\tilde q$ & $0.000005$ & $0.000047$ & $0.000107$ \\
\hline
$l\tilde l$ & $0.000015$ & $0.000149$ & $0.000175$ \\
\hline
$5\times q\tilde q$ & $0.000006$ & $0.000067$ & $0.000161$ \\
\hline

\end{tabular}
\caption{\label{tab2} Predicted event rates for bino--like LSP, for the same
  combinations of $E_{\rm min}, \ M_X$ and primary $X$ decay mode as in
  Table~1. We show results for the three different mSUGRA scenarios of
  \cite{bod1}, with first generation squark masses of about 370 GeV (D1), 580
  GeV (D2)  and 1,000 GeV (D3). The background is essentially the same as in
  Table~1.}
\end{center}
\end{table}


Table~2 shows event rates for bino--like neutralino. In this case the
scattering cross section depends strongly on the squark mass
\cite{bk,luis,mele}. We therefore show results for three different scenarios
introduced in ref.\cite{bod1}, with first generation squark masses near 370,
580 and 1,000 GeV, respectively. We see that the event rate remains below one
event per year and teraton in all cases. This result seems much less promising
than that of earlier studies \cite{bdhh2,luis}. However, our rates are
actually comparable to those of ref.\cite{luis}, once the differences in
treatment are taken into account. To begin with, we assume that the $X$
particles are distributed like Dark Matter, i.e. clump in our galaxy. Assuming
a uniform distribution throughout the universe, as done in ref.\cite{luis},
increases the neutralino flux by about one order of magnitude \cite{bdhh2}.
The reason is that such a uniform distribution suppresses the proton flux due
to the GZK effect. One therefore has to increase the normalization in order to
match the observed flux. A more or less uniform distribution of $X$ particles
could be achieved only if they are bound to cosmological defects, which
nowadays are quite tightly constrained by analyses of cosmic microwave
background anisotropies \cite{cmbstring}. Moreover, we quote events per year,
whereas ref.\cite{luis} finds about five events per lifetime of the
experiment, taken to be three years. Finally, ref.\cite{luis} applies a cut
(of $10^9$ GeV) on the total energy of the incident neutralino, whereas our
cut is on the visible energy.

We note that for $E_{\rm min} = 10^6$ GeV, the ten body decay mode and $X
\rightarrow l \tilde l$ decays now generally lead to similar event rates. The
reason is that very energetic bino--like neutralinos lose energy considerably
faster than higgsino--like neutralinos do: for rather light squarks the cross
sections are comparable, but the energy loss per scattering is much larger for
bino--like states, which produce a squark with $m_{\tilde q} \gg m_{\lsp}$,
than for higgsino--like states, which produce a heavier neutralino or chargino
very close in mass to the LSP. The $5 \times q \tilde q$ decay mode has a
larger flux of softer neutralinos, which suffers less from propagation
effects; for bino--like neutralinos this largely compensates the reduction of
the rate due to the fact that the cross section is smaller at smaller LSP
energy. However, if $E_{\rm vis} > 10^9$ GeV is required, even the relatively
softer LSPs produced from the ten body decay mode will typically scatter
several times before reaching the detector. $X \rightarrow l \tilde l$ decays
are then again more favorable, due to its initially much larger flux of very
energetic neutralinos.

This brings us to a feature of our treatment which enhances the event rate
compared to the numbers of ref.\cite{luis}. In that analysis all neutralinos
were discarded that interact even once before reaching the detector. This is
not necessary, since this interaction will again yield a neutralino (from the
decay of the produced squark), with typically about half the energy of the
original LSP. Fig.~2 shows that this regeneration effect also leads to a much
milder dependence of the final event rate on the cross section, and hence on
the squark mass, than found in ref.\cite{luis}. Increasing the squark mass
reduces the cross section, and hence the event rate for given flux. However,
it also reduces the effect of neutralino propagation through the Earth,
i.e. it increases the flux. These two effects obviously tend to cancel. As a
result the event rate as function of $m_{\tilde q}$ shows a rather broad
maximum, the location of which depends on the cut on $E_{\rm vis}$. A lower
$E_{\rm vis}$ means that softer neutralinos can contribute. Since the cross
section increases with neutralino energy, softer neutralinos can tolerate
lighter squarks before suffering significant propagation losses. As a result,
at smaller $E_{\rm min}$ the maximum rate occurs for smaller squark mass. This
effect is less pronounced for primary $X \rightarrow l \tilde l$ decays, since
in this case the incident neutralino spectrum is in any case rather hard,
even if no cut on $E_{\rm vis}$ is applied.

\begin{figure}[h!] 
\begin{center}
\includegraphics[width=14cm]{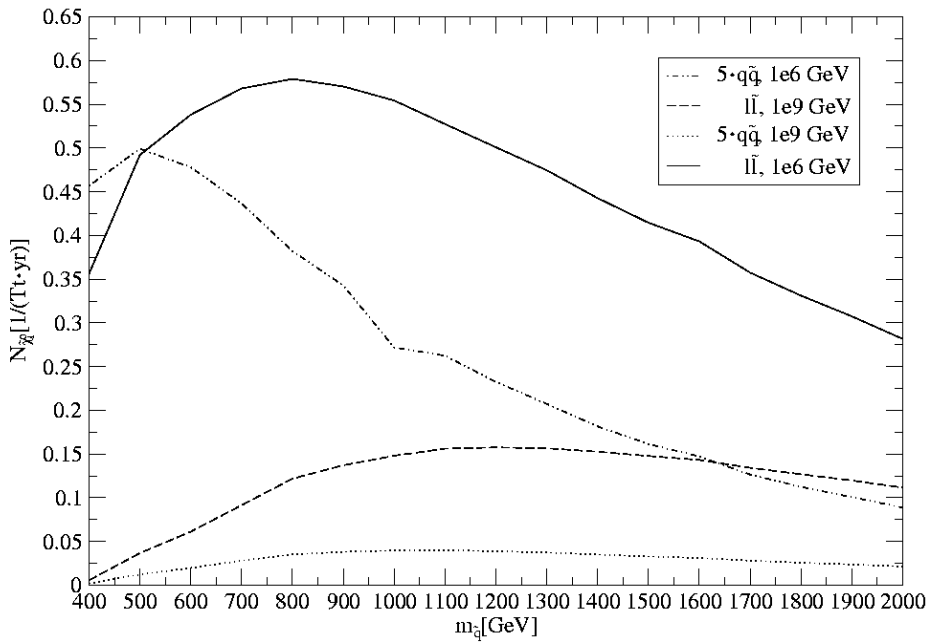}
\caption{Expected event rate due to bino--like neutralinos as function of the
  first generation squark mass, for two different primary $X$ decay modes and
  two choices of the minimal visible energy $E_{\rm min}$. See the text for
  further details.}
\label{fig2}
\end{center}
\end{figure}

\section{Summary and Conclusions}

In this paper we have calculated signal rates for the detection of very
energetic neutralinos, as predicted by ``top--down'' models for the observed
cosmic ray events at the highest energies. We use up--to--date calculations of
the neutralino flux generated at the location of the decay of the superheavy
particles, and of the effects due to propagation of the neutralinos through
the Earth. We also for the first time treat the case of higgsino--like
neutralino.

We conservatively assume that the progenitor ``$X$ particles'' are distributed
like Dark Matter, in which case most sources are ``local'', i.e. effects of
propagation through the interstellar or intergalactic medium are negligible.
We then find detectable event rates in teraton scale experiments with duty
cycle of $\sim 10\%$, typical for experiments based on optical methods, only
if the following conditions are satisfied: the lightest neutralino must be a
higgsino, rather than a bino; $M_X$ must be rather close to its lower bound of
$\sim 10^{12}$ GeV; and the experiment must either be able to detect upgoing
events with visible energy not much above $10^6$ GeV, or most $X$ particles
undergo two--body decays involving at least one slepton and no strongly
interacting (s)particle. The good news is that in all cases we studied the
signal is at least several times larger than the $\nu_\tau$ induced
background, computed in the same $X$ decay model. If $M_X$ is near $10^{16}$
GeV and the LSP is higgsino--like, or $M_X \sim 10^{12}$ GeV and the LSP is
bino--like, one will need ${\cal O}(100)$ Tt$\cdot$yr to collect a respectable
event rate. In the worst case, with a bino--like LSP, $M_X \sim 10^{16}$ GeV
and a threshold of the visible energy near $10^9$ GeV, one would observe less
than one event per year even if one monitored the entire surface of the Earth!
These numbers improve by about one order of magnitude if $X$ particles are
distributed more or less uniformly throughout the universe; this might be
expected if they are confined to cosmic strings or similar topological
defects. Recall, however, that scenarios with cosmic strings are constrained
by observations of cosmic microwave anisotropies.

These numbers only include interactions of neutralinos with nuclei. It has
been claimed in Ref.\cite{mele} that bino--like LSPs should lead to a
detectable signal in Gt class experiments (like IceCube \cite{ic}) through
resonant production of sleptons. However, they estimate the rates assuming a
neutralino flux close to the upper bound on the neutrino flux; the kind of
model we investigate here yields fluxes that are several orders of magnitude
smaller than this. Moreover, the visible energy in such events is relatively
small, since only the decay of the produced slepton contributes. At the
relevant energies the Earth does not filter tau neutrinos very well; so even
if one concentrates on upgoing events, the background in potentially realistic
$X$ decay models is several orders of magnitude larger than the signal.

Our overall conclusion is that next generation experiments, with effective
target masses in the Tt range, would have to be lucky to observe a signal from
neutralinos of ``top--down'' origin. Experiments with a relatively low energy
threshold would stand a much better chance than those with high threshold.
Unfortunately there are many reasonable $X$ decay scenarios where the
neutralino flux will remain invisible to such experiments. The goal of finding
an experimentum crucis for top--down models may therefore remain elusive.

\subsection*{Acknowledgments}
This work was partially supported by the Marie Curie Training Research Network
``UniverseNet'' under contract no.  MRTN-CT-2006-035863.

\end{document}